\documentclass[11pt]{article}
\usepackage{amsfonts, amssymb, amsthm}
\newfont{\extra}{msbm10 scaled\magstep1}

\newcommand{\extr}[1]{\mbox{\extra #1}}

\newcommand{\sect}[1]{\setcounter{equation}{0}\section{#1}}
\newcommand{\subsect}[1]{\subsection{#1}}

\renewcommand{\theequation} {\arabic{section}.\arabic{equation}}

\def\be{\begin{equation}}
\def\ee{\end{equation}}
\def\bea{\begin{eqnarray}}
\def\eea{\end{eqnarray}}

\def\b{\beta}

\def\k{\kappa}

\def\R{\extr R}

\begin{document}
\begin{center}{ \LARGE \bf
Anyons, group theory and planar physics
}
\end{center}
\vskip0.25cm

\begin{center}
J. Negro~$^1$, M. A. del Olmo~$^1$ and J. Tosiek~$^2$
\vskip0.25cm

$^1${\it Departamento de  F\'{\i}sica Te\'orica,\\
Universidad de  Valladolid,
 E-47011, Valladolid,  Spain} 
\vskip0.15cm

$^2${ \it 
Institute of Physics, Technical University of  L\'{o}d\'{z},\\ 
W\'{o}lcza\'{n}ska 219, 93-005 L\'{o}d\'{z}, Poland.}
\vskip0.15cm

E. mail: jnegro@fta.uva.es, olmo@fta.uva.es, tosiek@p.lodz.pl
\end{center}
\vskip1.5cm

\centerline{\today}
\vskip1.5cm

\begin{abstract}
Relativistic and nonrelativistic anyons are described in a unified formalism
by means of the coadjoint orbits of the symmetry groups in the free case as well as
when there is an interaction with a constant electromagnetic field.  To deal with 
interactions  we introduce
the extended Poincar\'e and Galilei Maxwell groups.
\end{abstract}

%%%%%%%%%%%%%%%%%%%%%%%%%%%%%%%%%%%%%%%%%%%%%%%%%%%%%%
%%%%%%%%%%%%%%%%%%%%%%%%%% INTRODUCTION %%%%%%%%%%%%%%%%%%%
\sect{Introduction}

The first works, of a rather theoretical character, devoted to
particles with an arbitrary spin and statistics in $(2{+}1)$ dimensions
go back to 1977 \cite{Leinaas}, but the real interest in
 physics behind the anyons started some years later when the fractional
quantum Hall effect was explained just in terms of anyons \cite{Laughlin}.
 
In the last years some works analyzing the role of the Galilei and
Poincar\'e groups in $(2{+}1)$ dimensions in the theory of anyons have
appeared in the literature \cite{ply91}-\cite{olmo05}.
Although the usual group theoretical considerations fit well when the anyons are
free, several difficulties appear when  the action of an external
electromagnetic field is considered \cite{ply05}. However,
recently \cite{Olmo04,Olmo04a} we have introduced two non-central  extensions 
of the Poincar\'e and Galilei groups by homogeneous and constant electromagnetic fields, 
called Maxwell groups \cite{negro90a,negro90b} that seem to provide an
appropriate  group theoretical framework for  anyons in the presence of constant
fields.  Our intention here is to adopt this viewpoint to present an unified
approach for these kind of interacting systems in 2-space + 1-time dimensions which
can also involve non-commuting coordinates. 
In this context we  mention a recent work \cite{horvathy04} where the
authors also obtain, in a different way, the extended Galilei-Maxwell group
(called by them ``enlarged Galilei group'').

The paper has been organized  as follows. 
In the first two Sections we revise the Poincar\'e and Galilei groups to recover 
the main features of the free relativistic and nonrelativistic anyons and fix the
notation.  As in all the  cases  presented along this paper,  we have made a
systematic use of the coadjoint orbit method that supply us with a canonical setup
of classical systems bearing enough symmetry in terms of the corresponding symmetry
group; in particular, for the Galilei group we have taken into account its double
central extension. In the next two sections we deal with interacting  anyons and
constant electromagnetic fields. First, in Section
\ref{Poincareanyonsfield} within a  relativistic frame, while the nonrelativistic
situation is considered in the following section, where we also discuss how to
perform the nonrelativistic limit. As we mentioned  before  the groups involved  in
our analysis are certain noncentral extensions of the Galilei and Poincar\'e groups
where the key point is that electromagnetic fields take part as dynamical 
objects.  Some conclusions and comments on the main differences with other
approaches end the paper.

We have also added two Appendices for the sake of completness. 
In a first Appendix we give a brief review about the symplectic
structures associated to a Lie group. 
The second Appendix 
supply with a classification of orbits for the space ${\cal
GM}^*(2{+}1)$ dual to the Lie algebra of the Galilei-Maxwell group.

%%%%%%%%%%%%%%%%%%%%%%%%%%% SECTION II %%%%%%%%%%%%%%%%%%%%%
%%%%%%%%%%%%%%%%%%%%%%%%%%%%%%%%%%%%%%%%%%%%%%%%%%%%%%

\sect{Anyons and the Poincar\'e group}\label{Poincareanyons}

The Poincar\'e group  $P(2{+}1)$, is a 6-dimensional (6-D) Lie group of
transformations of the \mbox{$(2{+}1)$--D} Minkowski spacetime provided
with the metric tensor $g_{i j}={\rm diag}\;(1,-1,-1)$. 
Two spatial cartesian axes will be denoted by $X_1$ and $X_2$.

Our (2{+}1)--D system may be seen as embedded in the
$(3{+}1)$--D Minkowski spacetime equipped with the metric tensor $g_{i
j}={\rm diag}\;(1,-1,-1,-1)$. Then, the third spatial axis perpendicular to
the $X_1X_2$-plane  will be denoted by $X_3$.  We will  make use
of this embedding; for example a rotation on the plane $X_1X_2$ may be
considered as the rotation around $X_3$. In that case vectors on the
spatial plane are thought to be $3$--D objects.

Each element of  $P(2{+}1)$ is
parametrized by a pair $(a,\Lambda)$, where 
$a =(b,a_1,a_2)$ represents a time
($b$) and space ($a_1,a_2$) translation, and $\Lambda$  a Lorentz
transformation. The element $\Lambda$ can be factorized as
$\Lambda=\Lambda(\chi,\vec{n})
\Lambda(R_\phi)$, with
$\Lambda(\chi,\vec{n})$  being a boost of rapidity $\chi$ in the
direction of the unit planar vector $\vec{n}$ and $\Lambda(R_\phi)$
denoting a rotation of angle
$\phi$ around the axis $X_3$.

The Lie algebra ${\cal P}(2{+}1)$ of the Poincar\'e group 
$P(2{+}1)$ is spanned by the basis 
$\{  P_0\equiv H, P_1,P_2, K_1,K_2, J \}$, which are the
infinitesimal generators of time and space 
translations, boosts transformations along axes $X_1,X_2$, and
$X_3$-rotation, respectively.
The nonvanishing commutators  are
\begin{equation}\label{commutationsPoincare}
\begin{array}{llllll}
&[H,K_i]=- P_i,&\qquad [P_i, K_i]= - H,&\qquad [ P_i,J]= -
\varepsilon _{ij} P_j,
\\[0.25cm]
&[K_i, K_j]=- \varepsilon_{ij} J,&\qquad [ K_i,J]= - \varepsilon
_{ij} K_j,&\qquad i,j=1,2,
\end{array}\end{equation}
where  $\varepsilon _{ij}$ denotes the 2--D completely skewsymmetric
tensor.
%%%%%%%%%%%%%%%%%%%%%%%%%%
%%%%%%%%%%%%%%%%%%%%%%%%%%

\subsect{Coadjoint orbits}

Let $(h,p_1,p_2,k_1,k_2,j)$ be the coordinates of an arbitrary
point of ${\cal P}^*(2{+}1)$, the dual space of ${\cal P}(2{+}1)$,
in a basis dual to $\{ H, P_1,P_2, K_1,K_2, J \}$. The coadjoint
action of $P(2{+}1)$ on ${\cal P}^*(2{+}1)$ is given by
\cite{martinarista}
\[
g: (h,p_1,p_2,k_1,k_2,j) \to (h',p_1',p_2',k_1',k_2',j'), \qquad
g=(b,a_1,a_2,\Lambda(\chi,\vec{n}),\,
\Lambda(R_\phi)),
\]
\begin{equation}
\begin{array}{lll}\label{coadjointaction}
h' &=&\cosh\chi\;h -\sinh \chi\; \vec{n}\cdot \vec{p}\;^\phi,\\[0.25cm]
\vec{p} \,' &=& \vec{p}\;^\phi  -\sinh \chi\;h\; \vec{n}
+ (\cosh\chi -1) (\vec{n}\cdot \vec{p}\;^\phi)\; \vec{n},\\[0.25cm]
\vec{k}' &=& \vec{k}\;^\phi  +\sinh \chi\;\vec{j}
\times \vec{n}\;^{\pi/2}
-(\cosh\chi -1) (\vec{n}\;^{\pi/2}\cdot \vec{k}\;^\phi)\; \vec{n}\;^{\pi/2}
\\[0.15cm]
 & &\quad +\; b\;[\vec{p}\;^\phi -\sinh \chi\;h\; \vec{n}+
 (\cosh\chi -1) (\vec{n}\cdot \vec{p}\;^\phi)\; \vec{n}]\\[0.15cm]
 & &\quad +\; \vec{a}\;[\cosh\chi\;h -\sinh \chi\; \vec{n}\cdot
\vec{p}\;^\phi], \\[0.25cm]
\vec{j}' &=& \cosh\chi\;\vec{j} + \sinh \chi\; \vec{n}\times
\vec{k}\;^\phi\\[0.15cm]
 & &\quad +\; \vec{a}\times [\vec{p}\;^\phi - \sinh \chi\;h\; \vec{n} +
(\cosh\chi -1) (\vec{n}\cdot \vec{p}\;^\phi)\; \vec{n}] ,
\end{array}\end{equation}
where we have used the following notation: 
\be
\label{licho0} 
\vec{n}=(n_1,n_2,0),\; \;\; \vec{p}=(p_1,p_2,0),\; \;\;
\vec{k}=(k_1,k_2,0),\; \; \; \vec{a}=(a_1,a_2,0),\; \;\; 
\vec{j}=(0,0,j),\ee 
and
$\vec{x}\;^\phi$ stands for the rotation of  a vector $\vec{x}$
 around the axis $X_3$ by an angle $\phi$.

The invariants of the coadjoint action (\ref{coadjointaction}) are
\begin{equation}\label{invariantsPoincare}
C_1=\rho^2 {=} g^{\mu\nu}\rho_\mu \rho_\nu =
h^2-p_1^2-p_2^2,\qquad C_2=h\; \vec{j} +\vec{p}\times \vec{k},
\end{equation}
where ${\rho}=(h,p_1,p_2)$ and $g_{\mu\nu}={\rm diag}(1,-1,-1)$.  

The invariant $C_2$ is in fact a $3$--D vector,
but its  only non-zero component is the third one, equal to 
$\;h j + p_1k_2 - k_1p_2$. It is a lower dimensional version of the
Pauli-Lubanski four-vector.  Recall that in $(3+1)$--D Minkowski spacetime the
Pauli-Lubanski vector $w$ takes the form 
\be 
 {w} {=}(w^0,{\bf w})=
({\bf j}\cdot {\bf p},h \; {\bf j} +{\bf p}\times {\bf k}), 
\ee
where now the involved vectors are generic 3--D.
The scalar ${w}^2$ is invariant under the $(3{+}1)$--Poincar\'e group
action.  In $(2{+}1)$ dimensions the Pauli-Lubanski vector reduces 
to the expression (\ref{invariantsPoincare}) of $C_2$.

 The classification of the coadjoint orbits was  published in
\cite{martinarista,binegar}. There are orbits of dimension 4, 2 and 0
(points).  The  $4$--D orbits are divided in three classes:
\begin{itemize}

\item
 $C_1 >
0 \;\;\;$ relativistic particles of a mass $\sqrt{C_1}$,

\item
 $C_1<0  \;\;\;$
tachyons,

\item
$C_1 =0, \; \vec{p} \neq 0  \;\;\; $ massless particles.

\end{itemize}

We will consider  the strata of orbits with $C_1 > 0$.
Rewriting $C_1 =m^2$,
where $m$ is the rest mass of the particle, 
the invariant $C_1$ leads to the equation of a hyperboloid of two
sheets  $h^2-p_1^2- p_2^2-m^2=0$. We will restrict to that one of
positive energy $h > 0$, denoted $H^+_m$, as usual.
For the second invariant, we rewrite $C_2= m\, \vec{s}$,  
where 
\be
\label{spin}
 \vec{s} = \frac{h\; \vec{j} +\vec{p}\times
 \vec{k}}{\sqrt{h^2-p_1^2-p_2^2}}\ .
\ee
The nontrivial component of $\vec{s}$ is the spin $s$ of the system. 

%%%%%%%%%%%%%%%%%%%%%%%%%
%%%%%%%%%%%%%%%%%%%%%%%%%

\subsection{Symplectic structure}

The two independent invariants $m$ and $s$ fix, in the way presented above,
a coadjoint orbit ${\cal O}^+_{m,s}$ that constitutes a $4$--D
differentiable submanifold of ${\cal P}^*(2{+}1)$.  Moreover,
we can cover  ${\cal O}_{m,s}^+$  with one  chart $({\cal
O}^+_{m,s},\varphi)$ using as  coordinates
 $(p_1,p_2,k_1,k_2).$
Indeed, the Jacobian of the transformation
\[
(h,p_1,p_2,k_1,k_2,j) \rightarrow (C_1,C_2,p_1,p_2,k_1,k_2)
\]
being $(2 h^2)^{-1}$, it is always
positive on the sheet $H^+_m$.
On the orbits ${\cal O}_{m,s}^+$ there is a natural Poisson structure
 (\ref{lambda}) ---see Appendix A--- given by
\begin{equation}
\Lambda={h}\; \frac{\partial}{\partial k_i}\wedge
\frac{\partial}{\partial p_i}  - {j}\;\frac{\partial}{\partial k_1}
\wedge \frac{\partial}{\partial k_2}\ .
\end{equation}
The symplectic form related with the tensor $\Lambda$ is
\begin{equation}
\omega=-\frac{1}{h}\; dk_i\wedge dp_i - \frac{j}{h^2}\; dk_1\wedge dk_2\,  ,
\end{equation}
where, according to (\ref{invariantsPoincare}) 
\be
\label{licho2} 
h=+\sqrt{\vec{p}\;^2+m^2}, \qquad j=\frac{m
s - p_1 k_2  + p_2 k_1 }{\sqrt{\vec{p}\;^2+m^2} }\ . \ee 
The coordinates, $p_i,\; k_i,\; i=1,2$ are not canonical since their Poisson
brackets are
\begin{equation}\label{poissonrellibres}
\{k_1,k_2\}= -{j} ,\qquad\{p_1,p_2\}= 0,\qquad
\{k_i,p_j\}= {h}\;\delta_{ij} .
\end{equation}
The equations of the time evolution obtained from the law of motion
(\ref{motionlow}) with the Hamiltonian (\ref{licho2}) are 
\be 
\label{pierwsze} \dot{p}_i=0,   \qquad \dot{k}_i=p_i ,\quad i=1,2. \ee
They look like the equations of motion of a {nonrelativistic } 
free particle. 

We find a set of canonical coordinates $\{ \vec{p},\vec{q}\}$,  where
\be
\label{licho3}
 \vec{{q}}=  \frac{\vec{k}}{h}- \frac{\vec{p}\times
\vec{s}}{h(m+h)}
\ee 
and the expression for the angular momentum becomes 
\be\label{licho21} 
j=\frac{m s }{h}+ \vec{q}\times  \vec{p}+ \frac{m s}{h(m+h)}\vec{p}\;^2. 
\ee 
Now, if we identify  ${q}_i$ as  `position coordinates', the equations of motion
are the well-known  relations
\be
\label{drugie} \dot{p}_i=0,  \qquad
\dot{q}_i=\frac{p_i}{h}, \qquad i=1,2. \ee 
A detailed analysis of the different coordinate systems for anyons can be found in 
Ref.\ \cite{mikhail96}.

%%%%%%%%%%%%%%%%%%%%%%%%%
%%%%%%%%%%%%%%%%%%%%%%%%%
\subsection{Irreducible unitary representations}

In quantum mechanics the coadjoint orbits of a Lie group  allow us to
define the irreducible unitary representations (IUR) associated to quantum
elementary physical systems having such a symmetry group. Thus, the
IUR's of
${P}(2{+}1)$ associated to the stratum of orbits ${\cal O}_{m,s}^+$ are
\begin{equation}
[U_{m,s}(a,\Lambda)] \psi(p)=e^{i p\cdot a} e^{i s\,\theta(p,\Lambda)}
\psi(\Lambda^{-1}p),
\end{equation}
where $s$ is the quantum number labelling a representation
of  $SO(2)$ and $\theta(p,\Lambda)$ is the Wigner angle, which is
determined by  the little group of a point of the orbit.
More explicitly,
choosing the point
$p_m=(m,0,0)$, whose isotropy group is $SO(2)$, and
the boost elements $\Lambda_{p\to p'}\in SO(2,1)$ transforming the
point $p$ into $p'$, then
\begin{equation}
\theta(p,\Lambda)=\Lambda^{-1}_{p_m\to \Lambda(p)}\;\Lambda\;
\Lambda_{p_m\to p} .
\end{equation}
 The functions
$\psi(p)$ belong to the Hilbert space ${\cal H}={\cal L}^2(H^+_m,
d\mu(p))$, being
$d\mu(p)$ the
$SO(2,1)$--invariant measure in $H^+_m$. 

The differential realization of the generators for this representation is
\begin{equation}
\hat H=h, \quad \hat P_i=p_i, \quad
\hat K_j=i(h \partial_{p_j}+ p_j \partial_h)
+ \frac{\varepsilon^{jk}  p_k}{2 m} s,
\quad
\hat J=i (p_2\partial_{p_1}-p_1\partial_{p_2}) +\frac h m s .
\end{equation}
The two Casimirs corresponding to the
invariants (\ref{invariantsPoincare}), $C_1=\hat P^2$ and
$C_2= \hat H\cdot \hat J +\vec{\hat P}\times \vec{\hat K}$, give the following
equations
\begin{equation}\label{difeq}
(p^2-m^2)\; \psi(p)=0, \qquad
(h\; \hat J +\vec{\hat P}\times \vec{\hat K}  -m s )\;\psi(p)=0 .
\end{equation}
The first one corresponds to the mass shell condition which gives rise to
the Klein-Gordon equation. The second one is the Pauli--Lubanski equation
describing the spin of the particle. In two dimensions
the unitarity of the realizations does not impose restrictions on
the values of $s$, thus allowing for
the existence of anyons. In this way we   easily recover  results of 
Ref.~\cite{jackiw91}.

%%%%%%%%%%%%%%%%%%%%%%%%%%% SECTION IV %%%%%%%%%%%%%%%%%%%%%%%%%%%%%%%
%%%%%%%%%%%%%%%%%%%%%%%%%%%%%%%%%%%%%%%%%%%%%%%%%%%%%%%%%%%%%%%%
\sect{Nonrelativistic anyons and the Galilei group}\label{galileianyons}

In the nonrelativistic case we have  to deal  with the Galilei
group ${G}(2{+}1)$ in $(2{+}1)$--D, which can be seen as a contraction of
the Poincar\'e group $P(2{+}1)$. The commutation rules of its Lie
algebra ${\cal G}(2{+}1)$ are those of Poincar\'e 
(\ref{commutationsPoincare}) except that now
\be
 [K_1,K_2]=0 , \qquad
[K_i,P_j]=0, \qquad i,j=1,2. 
\ee
The  algebra ${\cal G}(2{+}1)$ admits a 2--D central extension 
$\bar{\cal G}(2{+}1)$
characterized by the new commutators \cite{levyleblond,ballesteros}
\begin{equation}\label{commutationsgalilei}
[ P_i,K_j]=- \delta _{ij} M,\qquad [K_1, K_2]={\cal K} ,
\end{equation}
 where $M$ and ${\cal K}$ are central generators, i.e., $[M,\cdot]=
 [{\cal K},\cdot]=0$ for any element of ${\cal G}(2{+}1)$.

Both extensions can also be obtained by a contraction from
the Poincar\'e group~\cite{ballesteros,arratia97}. It is enough to
consider  the direct product $\tilde P(2{+}1)=\R^2\otimes
P(2{+}1)$. Obviously, at the level of the Lie algebra we have 
$\tilde {\cal P}(2{+}1)=\R^2\oplus {\cal P}(2{+}1)$. Hence, a basis is
constituted by the generators of $\R^2$ ($M,{\cal K}$) plus the known
generators of the  Poincar\'e algebra  ($H,P_1,P_2, K_1,K_2, J$).
Let us consider a new basis given by
\be\label{basenueva1} 
 M'=M,\quad  {\cal K}'={\cal K},\quad  H'=H-M,\quad 
 P'_i=P_i, \quad  K'_i=K_i,\quad  J'=J+{\cal K},\qquad
i=1,2. 
\ee 
The nonvanishing commutators of $\tilde {\cal P}(2{+}1)$ in this new basis
are
\be\begin{array}{llllll} \label{tildecommutators} 
&[
H',  K'_i]= -  P'_i,&\qquad &[  P'_i, K'_i]=-
 H' -  M', & \qquad 
&[  P'_i, J']= -
\varepsilon _{ij} P'_j,
 \\[0.25cm]
&[  K'_i,  K'_j]=- \varepsilon_{ij} (  J' -{\cal K}'),
 &\qquad &[
K'_i,  J']= - \varepsilon _{ij}  K'_j, &\qquad
&i,j=1,2.
\end{array}
\ee
Now in order to perform the contraction we define an appropriate rescaled basis
\be\label{basenueva2} 
 M''=\epsilon^2  M',\quad 
{\cal K}''=\epsilon^2 {\cal K}',\quad  H''=  H',\quad
 P''_i=\epsilon   P'_i, \quad   K''_i=\epsilon  
K'_i,\quad   J''=  J ',
\ee 
where $\epsilon$  is a
fixed real positive number.
The nonvanishing Lie commutators  are now 
\[
\begin{array}{llllll}
\label{tildeprimacommutators} 
&[ H'',  K''_i]= -  
P''_i,&\  &[  P'_i,  K'_i ]= -\epsilon^2( H'' 
+\frac{1}{\epsilon^2} M''),&\  & [ P''_i,
J'']=- \varepsilon _{ij}  P''_j,
\\[0.25cm]
& [ K''_i,   K''_j]=- \varepsilon_{ij} \epsilon^2( 
J'' -\frac{1}{\epsilon^2}  {\cal K}''), &\  &  [  
K''_i,  J'']= - \varepsilon _{ij}   K''_j, &\ 
&i,j=1,2.
\end{array}\]
In the limit $\epsilon \rightarrow 0$ we recover the Lie
commutators of the extended  algebra $\bar{\cal G}(2{+}1)$.

To give a physical interpretation of the contraction procedure we
identify the contraction parameter $\epsilon$ with the inverse of
the light speed ($\epsilon=1/c$). From a
cohomological point of view a  change of the basis defined by
relations  (\ref{basenueva1})-(\ref{basenueva2}) corresponds to introducing a
trivial two-cocycle on the Poincar\'e group. After the contraction $\epsilon
\rightarrow 0$ this trivial two-cocyle becomes a non-trivial one of the Galilei
group \cite{adolfo}.
%%%%%%%%%%%%%%%%%%%%%%%%%%
%%%%%%%%%%%%%%%%%%%%%%%%%%
\subsect{Coadjoint orbits}

By $\bar{\cal G}^*(2{+}1)$ we will denote the space dual to the algebra
$\bar{\cal G}(2{+}1).$  Each vector belonging to $\bar{\cal G}^*(2{+}1)$
is characterized by eight components $(m,\k , h,p_1,p_2,k_1,k_2,j)$ in
the basis dual to $(M,{\cal K},H,P_1,P_2,K_1,K_2,J)$ of $\bar{\cal
G}(2{+}1)$. 

Let us denote by $g=(\theta, \eta, b,\vec a,\vec v,R_\phi)$
the elements of $\bar{ G}(2{+}1)$, with a convention similar to that of Poincar\'e
${\cal P}(2{+}1)$ except that, $\vec v$ stands for the Galilean boosts,
and $(\theta,\eta)$ parametrize the group elements generated by $(M,{\cal K})$.
The coadjoint  action of $g\in \tilde G(2{+}1)$
on the dual space $\bar {\cal G}^*(2{+}1)$ is given by \cite{ballesteros}
\begin{equation}\begin{array}{lll}\label{coadjointactiongalilei}
m' &=& m , \\[0.25cm]
{\k}' &=& {\k} , \\[0.25cm]
h' &=&h -  \vec{v}\cdot \vec{p}\;^\phi + \frac{1}{2}m
\vec{v}\;^2,\\[0.25cm]
\vec{p}\;' &=& \vec{p}\;^\phi  -m \vec{v} ,\\[0.25cm]
\vec{k}\;' &=& \vec{k}\;^\phi  + b\; \vec{p}\;^{\phi}
+ m\; (\vec{a}- b\;  \vec{v}) + \vec{v}\times \vec{\k},
\\[0.25cm] \vec{j}\;' &=& \vec{j} +  \vec{a}\times \vec{p}\;^\phi +
\vec{v}\times
\vec{k}\;^\phi -
\frac{1}{2} \;\k\; \vec{v}\;^2 - m \;\vec{a}\times \vec{v}  ,
\end{array}\end{equation}
where we have also used the notation
\[
 \vec{p}=(p_1,p_2,0), \;\; \vec{v}=(v_1,v_2,0), \;\;  
\vec{k}=(k_1,k_2,0), \; \;  \vec{a}=(a_1,a_2,0), \;\; 
 \vec{\k}=(0,0,\k), \;\;  \vec{j}=(0,0,j). 
\] 

The invariants of the coadjoint action
(\ref{coadjointactiongalilei}), besides $m$ and $\k$, are
\begin{equation}\label{galinvariants}
 C_1=\vec{p}\;^2 -2\; m\; h  ,\qquad C_2= m\;
\vec{j} -\vec{\k} \; h + \vec{p}\times \vec{k}.
\end{equation}
Note that the first one can be written as 
$ U= - {C_1}/{2m}$ and is interpreted as the internal
energy  of the physical system. As in the relativistic case we denote
$C_2=m \vec{s}$, but now 
\be\label{snorelativista}
\vec{s}\,{=}\,\vec{j}-\frac{\vec{\k} h}{m} + \frac{\vec{p}\: \times
\vec{k} }{m}.
\ee 
It is easy to derive the expressions (\ref{galinvariants}) from their
relativistic counterparts (\ref{invariantsPoincare}) following the
contraction procedure outlined above (see also Ref.~\cite{arratia97}). 
Obviously, expression  (\ref{snorelativista}) is the nonrelativistic limit of 
(\ref{spin}).

The classification of the coadjoint orbits (4--D, 2--D and 0--D) was presented in
\cite{ballesteros}.  The  relevant 4--D orbits  characterized
by the values
$\{m\neq 0, \k ,  U, \; s\}$ are  denoted by  ${\cal O}^{\k, U}_{m,s}$. 
In the following we will assume $\k\neq 0$, since the results for
$\k=0$ can be obtained directly. 
%%%%%%%%%%%%%%%%%%%%%%%%
%%%%%%%%%%%%%%%%%%%%%%%%
\subsection{Symplectic structure}

A set of coordinates adapted
to the orbit ${\cal O}^{\k,U}_{m,s}$ are
$(m,\k,U,s,p_1,p_2,x_1,x_2)$, where $x_i = k_i/m$.  Since the transformation
\[
(m,\k,h,p_1,p_2,k_1,k_2,j)\rightarrow (m,\k,U,s,p_1,p_2,x_1=k_1/m,x_2=k_2/m)
\]
has a nonzero Jacobian (as long as $m \neq 0$!), we can cover the whole
orbit 
${\cal O}^{\k,U}_{m,s}$ with one chart $({\cal O}^{\k,U}_{m,s},\varphi)$
using coordinates $(p_1,p_2,x_1,x_2)$.  
The induced Poisson tensor
$\Lambda$ on the orbit ${\cal O}^{\k, U}_{m,s}$ takes the form
\begin{equation}
\Lambda= \frac{\partial}{\partial x_i}\wedge
\frac{\partial}{\partial p_i}  +
\frac{\kappa}{m^2}\;\frac{\partial}{\partial x_1} \wedge
\frac{\partial}{\partial x_2}, \qquad i=1,2.
\end{equation}  
The inverse of $\Lambda$  gives the  symplectic form $\omega$ on 
${\cal O}^{\k, U}_{m,s}$,
\begin{equation}
\omega= dx_i\wedge dp_i + \frac{\k}{m^2} \;
dx_1\wedge dx_2, \qquad i=1,2.
\end{equation}
The coordinates $(\vec{p},\vec{x})$  are not canonical since their
Poisson brackets are
\begin{equation}
\{x_1,x_2\}= \frac{\k}{m^2} ,\qquad\{p_1,p_2\}= 0,\qquad
\{x_i,p_j\}=\delta_{ij} .
\end{equation}
Nevertheless, the Hamiltonian 
\be \label{ha} 
h= \frac{\vec{p}^{\:2}}{2m} + U
\ee 
has the usual form of a free nonrelativistic particle,
leading to the motion equations
\be
\label{rownaniadyn1}
\dot{\vec{p}}=0, \qquad  \dot{\vec{x}}= \frac{\vec{p}}{m}.
\ee
We see that the dynamics of the particle is independent of the parameter
$\k$  because  the Hamilton function (\ref{ha}) does not contain any
function of $\vec x .$
 Equations (\ref{rownaniadyn1})
  are  consistent with
 (\ref{coadjointactiongalilei}) if we assume that the parameter
 $b$ represents the time and the coadjoint
action gives the relation between the coordinates in
two different inertial frames. Indeed, let the
laboratory frame $\Sigma '$ and the (instantaneous) rest frame $\Sigma$ of
the particle be related by the element
$g=(t,\vec{a},-\vec{v},0)\in  G(2{+}1).$  Then, equations
(\ref{coadjointactiongalilei}) can be read as 
\be
\label{licho04}
\vec{p}_{\Sigma'}= m \vec{v} ,\qquad
 \vec{x}_{\Sigma'}=t \: \frac{\vec{p}_{\Sigma'}}{m}- 
\frac{\vec{p}_{\Sigma'}}{ m^2} \times \vec{\k} + \vec{a}. 
\ee
The angular momentum is given by
\be
\label{licho210} 
j= \vec{x}\times  \vec{p}+ \frac{\vec{\k}}{m}\left( \frac{\vec{p}^{\:2}}{2m} + U
\right) +s 
\ee 
which is the nonrelativistic limit of (\ref{licho21}).

A set of canonical coordinates 
$(\vec{p},\vec{q})$ can
be straightforwardly obtained from $(\vec{p},\vec{k})$ by  
\be\label{tilde}
{q}_i= x_i+ \frac{\varepsilon_{ij}\k p_j}{2m^2}.  
\ee
It is worth to consider (\ref{tilde}) as the nonrelativistic limit of the
corresponding ones (\ref{licho3}) for Poincar\'e.
The coadjoint action of the Galilei group $G(2{+}1)$  in these
coordinates is expressed by formulas
\begin{equation}\begin{array}{l}\label{coadjointactiongalilei1}
\vec{{p}}~' = \vec{{p}}\;^\phi  -m \vec{v} ,\\[0.25cm]
\vec{q}~' = \vec{q}\;^\phi  + \frac{b}{m}\;
\vec{{p}}\;^{\phi}
+  (\vec{a}- b\;  \vec{v}) + \frac{1}{2m}\vec{v}\times \vec{\k}.
\end{array}\end{equation}
Although $\k$ does not affect the dynamics, it
gives the contribution $\frac{1}{2m}\vec{v}\times \vec{\k}$ to
$\vec{q}.$
The time evolution of canonical variables is given by (cf.
(\ref{rownaniadyn1})) 
\be \dot{p}_i=0, \qquad
\dot{q}_i= {p}_i, \qquad i=1,2. 
\ee 
 Note that in the coordinates
$(\vec{p},\vec{q})$
 the angular momentum  takes the form
\begin{equation}
\vec{{j}}= \vec{q} \times \vec{p}
+\frac{\vec{\k}\,U}{m} + s ,
\end{equation}
where $\k$ gives rise to an extra term.
Similarly to (\ref{licho04}) equations
(\ref{coadjointactiongalilei1}) give now
\be
\label{licho4}
\vec{p}= m \vec{v} ,\qquad
 \vec{q}=t \: \frac{\vec{p}}{m}- 
\frac{\vec{p}}{2 m^2} \times \vec{\k} + a . 
\ee
The $\k$--term is the only one that remains without a  clear physical
interpretation in the free case \cite{ballesteros,brihaye}.

%%%%%%%%%%%%%%%%%%%%%%%%
%%%%%%%%%%%%%%%%%%%%%%%%
\subsection{Irreducible unitary representations}

The IUR of ${G}(2{+}1)$ associated to this
stratum of orbits are \cite{ballesteros}
\begin{equation}
[U_{m,s}^{\k,U}(g)] \psi(\vec{p})=e^{i ((\frac{1}{2m}\vec{p}\;^2+ U) b
- \vec{p}\cdot \vec{a})}
e^{i \k(\frac{1}{2m}\vec{v}\times \vec{p})}
e^{i (s+\kappa \frac U m)\phi}
\psi(R^{-1}(\phi)(\vec{p} - m \vec{v})),
\end{equation}
where $R (\phi)$ is a rotation of angle $\phi$. The carrier space of
the representation is the Hilbert space
${\cal H}={\cal L}^2(\R ^2)$.
Note that the differential realization of the generators in this
representation is
\begin{equation}
\hat P_j=p_j, \quad  \hat K_j=i m \partial _{p_j}
-\frac{\k}{2m}\varepsilon^{jk}  p_k, \quad
\hat H=\frac{1}{2m} \vec{p}\;^2 +U,  \quad
\hat J=i (p_2\partial_{p_1}-p_1\partial_{p_2}) +s+\k\, U .
\end{equation}
After a global change of phase, $U'(g)=\lambda (g) U(g)$, the IUR
$U_{m,s}^{\k,U}$ is shown to be equivalent to $U_{m,0}^{\k,0}$. 

It is worth mentioning that we can consider massless particles in the
Galilean framework.
In this case their 4-D coadjoint orbits are characterized by 
($m=0, \k\neq 0$) and the invariants $C_1=\vec{p}\;^2\neq 0$ and 
$C_2= \vec{p}\times \vec{k}-\k h$. If we compare $C_2$ with 
the relativistic Pauli-Lubanski operator for $m=0$ in (\ref{difeq}),
$(h\; \hat J +\vec{\hat P}\times \vec{\hat K}  -C_2 )\;\psi(p)=0$,
we see that $C_2$ gives the helicity of our system, while $J$ is now
replaced by $\k$. We remark that this case is not equivalent to that of
the orbit ($m=0, \k = 0$), and that in the present context the Hamiltonian
becomes linear in $\vec p$. However, one must be careful about the
interpretation since the coordinates $(\vec k, \vec p)$ are not canonical
(see also \cite{jackiw91}).

The IUR associated to the null--mass 
orbits are
\begin{equation} [U_{C_1,C_2}^{\k}(g)] \psi(\vec{w}, \theta)=
e^{i \k(\frac{1}{2}\vec{w}\times \vec{y})}
e^{i \vec{p}\cdot (\vec{a}- b \vec{y})}
e^{-i \frac{C_2}{2 \k}b}
\psi(\vec{y} - \vec{w},\theta -\phi),
\end{equation}
where $\vec{p}=\sqrt{C_1}(\cos \theta, \sin \theta)$ and $\psi \in {\cal
L}^2(\R ^2\times S^1)$. 

%%%%%%%%%%%%%%%%%%%%%%%%%%%%%%%SECTION V %%%%%%%%%%%%%%%%%%%%%%%%%%%%%%%%
%%%%%%%%%%%%%%%%%%%%%%%%%%%%%%%%%%%%%%%%%%%%%%%%%%%%%%%%%%%%%%%%%%%%%%%%%%

\sect{Relativistic anyons in an external electromagnetic field}
\label{Poincareanyonsfield}

Once revisited the description of free anyons,
in the next sections we will analyze the more interesting case when
charged particles move in a constant electromagnetic field. The presence
of external forces modifies the symmetry group of the system. This is the
reason why instead of the Poincar\'{e} $P(2{+}1)$ and  Galilei $G(2{+}1)$
groups, we will consider the so called  Poincar\'e--Maxwell
$PM(2{+}1)$ and Galilei--Maxwell $GM(2{+}1)$ groups
\cite{negro90a,negro90b}.

Let us start with the relativistic case.  The Poincar\'e--Maxwell
group $PM(2{+}1)$ is a 9--D Lie group with six infinitesimal generators
$\{H,P_1,P_2,K_1,K_2,J\}$,
corresponding to $P(2{+}1)$, plus three new
elements $\{B,E_1,E_2\}$ related to the electromagnetic field
\cite{negro90a}. It can be considered as a 3--D noncentral extension of
the Poincar\'e group.
The nonvanishing commutators for its Lie algebra, ${\cal PM}(2{+}1)$, are:
\begin{equation}
\begin{array}{lll}
[B,K_i]= \varepsilon_{ij}E_j, &  [E_i,K_j]= -\varepsilon_{ij} B, &
[E_i, J]= - \varepsilon_{ij} E_j, \\[0.25cm]
 [H,P_i]= E_i, & [H,K_i]= - P_i, & [P_i,P_j]= -
\varepsilon_{ij} B,  \\[0.25cm]
    [P_i,K_j]= - \delta_{i j} H, & [P_i,J]= - \varepsilon_{ij} P_j, &
[K_i,K_j]= -\varepsilon_{ij} J,  \\ [0.25cm]
 & [K_i,J]= -\varepsilon_{ij} K_j,    &\qquad i,j=1,2.
\end{array}
\end{equation}
%%%%%%%%%%%%%%%%%%%%%%%%%%%%%%%
%%%%%%%%%%%%%%%%%%%%%%%%%%%%%%%
\subsection{Coadjoint orbits}

We will denote by $(\b,\epsilon_1,\epsilon_2,h,p_1,p_2,k_1,k_2,j)$ the coordinates
fixing a point on ${\cal PM}^*(2{+}1)$ by means of the dual basis of
$(B,E_1,E_2,H,P_1,P_2,K_1,K_2,J)$. The general formula expressing
the coadjoint action of a group element
$g=(c,\vec d,b,\vec a,\Lambda(\chi,\vec n)\Lambda(R_\phi))\in PM(2{+}1)$
on
${\cal PM}^*(2{+}1)$  is given by 
\be\label{c}\begin{array}{l}
 \vec{\b}'=  (\cosh{\chi})\vec{\b}+(\sinh{\chi})\:({\vec n}\cdot{\vec \epsilon}^{\:\phi})\:
{\vec n}^{\:\pi/2}, \\[0.25cm]
 \vec{\epsilon }\:'= \vec{\epsilon } - (\sinh{\chi}) (\vec{\beta }\times {\vec
 n}^{\:\pi/2})
+(\cosh{\chi}-1) ({\vec n}^{\:\pi/2}\cdot{\vec \epsilon }^{\:\phi})\:
{\vec n}^{\:\pi/2}, \\[0.25cm]
 h'= (\cosh{\chi}) h - (\sinh{\chi})({\vec n}\cdot{\vec p}~^{\phi})
-\vec{a} \cdot \vec{\epsilon }\:', \\[0.25cm]
 \vec{p}\:'= \vec{p}~^{\phi}  -(\sinh{\chi}) h\: {\vec n}
+(\cosh{\chi}-1) ({\vec n}\cdot{\vec p}~^{\phi})\:{\vec n} + b \vec{\epsilon }\:'
- \vec{\beta }' \times \vec{a}, \\[0.25cm]
 \vec{k}\:'=\vec{k}^{\phi} - (\sinh{\chi}) \vec{j}\times {\vec n}^{\:\pi/2}
+(\cosh{\chi}-1) ({\vec n}^{\:\pi/2}\times{\vec k}~^{\phi})\: {\vec n}^{\:\pi/2}\\[0.25cm]
\hskip1cm + \: b \vec{p}~' -
\frac{1}{2} b^2 \vec{\epsilon }\:' + h'\vec{a}+ \frac{1}{2}\vec{a}^2 \vec{\epsilon }\:'
-   \vec{\beta }' \times \vec{d} - \vec{c}\times \vec{\epsilon }\:', \\[0.25cm]
 \vec{j}\:'= (\cosh{\chi})\vec{j} +(\sinh{\chi})\:({\vec n}\times{\vec k}^{\phi})
+ \vec{a}\times  \vec{p}~'
+ \frac{1}{2}\vec{a}^2 \vec{\beta }'+ \vec{d} \times \vec{\epsilon }\:',
\end{array}\ee
where the notation used is the same as in (\ref{licho0}) with the additional
vectors
\[
\vec{\beta }=(0,0,\beta ), \qquad
\vec{\epsilon }=(\epsilon _1,\epsilon _2,0), \qquad
\vec{c}=(0,0,c) ,\qquad
\vec{d}=(d_1,d_2,0).
\] 
The parameters $c,d_1$ and $d_2$ describe the group elements generated by 
$B, E_1$ and $E_2$, respectively.
Notice that $(h,p_1,p_2)$ represents the 3-D
energy-momentum vector covariant under (2+1) Lorentz transformations.

The invariants under the coadjoint
action (\ref{c}) are
\be\label{pomoc}\begin{array}{l}
 C_0=    \vec{\epsilon }\;^2 - \vec{\beta }\;^2 ,\\[0.25cm]
 C_1=h^2-\vec{p}^{\:2} -2 ( 
\vec{k} \cdot \vec{\epsilon }-\vec{j} \cdot \vec{\beta }\, ),  \\[0.25cm]
 C_2=h \,\vec{\beta } + \vec{p} \times \vec{\epsilon }.
\end{array}\ee
The first one is, of course, the invariant of the
electromagnetic field under Lorentz transformations, $C_0 =
-F_{\mu\nu}F^{\mu\nu}$, where $F_{0i}= \epsilon_i$ and
$F_{12}=\beta$.  If $C_0>0$, $C_0<0$ or $C_0=0$
we say that the electromagnetic field (or the orbit) is of electric, magnetic or
perpendicular type, respectively. 

The second invariant $C_1$ describes the interaction: it
includes the electric coupling term $\vec{k} \cdot \vec{\epsilon }$, and
the
coupling of angular momentum and magnetic field, $\vec{j} \cdot \vec{\beta
}$.
The last invariant $C_2$ from (\ref{pomoc}) admits a covariant expression
$C_2=-\varepsilon^{\mu\nu\sigma}p_{\mu}F_{\nu\sigma}$. It has not an immediate
interpretation, but its appearance is a consequence of the symmetries of the
system.

Since we have three independent invariants the maximal dimension of the
coadjoint orbits is $6$. In this work we will be concerned only with
this kind of orbits, henceforth denoted ${\cal O}^{C_0}_{C_1\:C_2}$.

%%%%%%%%%%%%%%%%%%%%%%%%%%%%%%%%
%%%%%%%%%%%%%%%%%%%%%%%%%%%%%%%%
\subsection{Symplectic structure}

A suitable chart of coordinates for the points of
the 9-D differentiable manifold ${\cal PM}^*(2{+}1)$ is given by
$(C_0,C_1,C_2,\epsilon _1,\epsilon _2,p_1,p_2,k_1,k_2)$. Each 6--D orbit
${\cal O}^{C_0}_{C_1\:C_2}$ can be covered with just one chart  $({\cal
O}^{C_0}_{C_1\:C_2},\varphi)$ with coordinates 
$(\epsilon _1,\,\epsilon _2,\,p_1,\,p_2,\,k_1,\,k_2)$.
Indeed, from (\ref{pomoc}) we find that the Jacobian of the
transformation
\[
(\beta ,\,h,\,j,\,\epsilon _1,\,\epsilon _2,\,p_1,\,p_2,\,k_1,\,k_2)
\longrightarrow (C_0,\,C_1,\,C_2,\,\epsilon _1,\,\epsilon _2,\,p_1,\,p_2,\,k_1,k_2)
\]
equals
$4 \beta ^3$. A singularity appears if $\beta =0$, but
we shall deal here only with orbits of magnetic type ($\beta \neq 0$).

In the chart $({\cal O}^{C_0}_{C_1\:C_2},\varphi)$ the Poisson
tensor $\Lambda$ takes the form 
\be \label{lam1}
\Lambda= -\beta  \frac{\partial}{\partial \epsilon _1} \wedge
\frac{\partial}{\partial k_2} +\beta  \frac{\partial}{\partial \epsilon _2}
\wedge \frac{\partial}{\partial k_1}
-\beta  \frac{\partial}{\partial p_1} \wedge \frac{\partial}{\partial p_2}
-h  \frac{\partial}{\partial p_1} \wedge \frac{\partial}{\partial k_1}
-h \frac{\partial}{\partial p_2} \wedge \frac{\partial}{\partial k_2}
-j  \frac{\partial}{\partial k_1} \wedge \frac{\partial}{\partial k_2},
\ee
where $\beta ,h,j$ are functions of the coordinates obtained
from  relations (\ref{pomoc}).
To make easy further comparisons with the nonrelativistic case we will
represent it as a matrix (using the previous order of the coordinates)
\begin{equation}
\label{macierzer}
\Lambda^{ij} = \left(
\begin{array}{cccccc}
0&0&0&0&0& -\beta  \\
0&0&0&0& \beta  &0 \\
0&0&0&-\beta  &-h & 0 \\
0&0&\beta &0&0&-h \\
0&-\beta &h&0&0&-j \\
\beta &0&0&h&j&0
\end{array}
\right)
\end{equation}
whose determinant  is $\beta ^6.$  The symplectic two-form is
\be
 \omega= \frac{h^2 -j \beta }{\beta ^3} d\epsilon_1 \wedge d\epsilon_2 +
\frac{h}{\beta^2} d\epsilon_1 \wedge dp_1 +\frac{1}{\beta } d\epsilon_1 \wedge dk_2
\frac{h}{\beta^2} d\epsilon_2 \wedge dp_2 - \frac{1}{\beta } d\epsilon_2 \wedge dk_1 +
\frac{1}{\beta } dp_1
\wedge dp_2.
\ee
Such a symplectic form is not canonical: the nonvanishing Poisson brackets
are
\be\label{b}
\{\epsilon_i,k_l\}= -\varepsilon_{il}\, \beta , \quad
\{p_i,p_l\}=\varepsilon_{il}\,  \beta ,  \quad
\{p_i,k_l\}= - \delta_{il}h, \quad \{k_i,k_l\}=- \varepsilon_{il}\, j, \qquad i,l=1,2.
\ee

In order to  find the equations of motion  we need to know the Hamiltonian
of the
system. From the invariant  $C_2$ (see (\ref{pomoc})) we get
\be
\label{f}
h = \frac{C_2}{\beta } + \frac{\vec{\epsilon } \times \vec{p}}{\beta },
\ee
where from (\ref{pomoc}) we can write
$ \beta =  \sqrt{\vec{\epsilon}\;^2 - C_0} $. Thus, we obtain 
\be \label{g1}
 \dot{\epsilon _i}= 0,  \qquad  \dot{p_i}= - \epsilon _i, \qquad
\dot{k}_i=p_i, \qquad i=1,2 .
\ee
Equations
(\ref{g1})  are extremely simple. The first one says that the fields are
constant. The others look like the
equations of motion for a  nonrelativistic
 particle with a unit negative charge inside
an electric field. Comparing the time evolution from these formulae with
the transformation rules (\ref{c}) we can identify the time $t$ with the
parameter $b$. This is natural because $b$ is associated to the
Hamiltonian generator $H$.

Let us remark that the above relations are quite different from the
usual ones derived from the standard Hamiltonian
formalism. In that approach the Hamilton function is
\be \label{f1} 
H =
\sqrt{m^2 + (\vec{\pi} + \frac{1}{2}\vec{\beta } \times \vec{r})^2} -
\vec{\epsilon } \cdot \vec{r}, 
\ee 
where $m$ denotes the mass of the
particle and $\vec{r}$ its vector of position, being 
$(\vec{\pi},\vec{r})$ canonically conjugated
variables. Then, the equations of
motion are  
\be\label{f3} 
\dot{\vec{\pi}}= \vec{\epsilon } -
\frac{1}{2}\dot{\vec{r}} \times \vec{\beta }, \qquad \dot{\vec{r}}=
\frac{1}{\cal E}({\vec{\pi}+ \frac{1}{2} \vec{\beta } \times \vec{r}}\:),
\ee 
with ${\cal E}=H+\vec{\epsilon }\cdot\vec{r}$  the energy of
the system.  
But, the  ``group'' coordinates $\epsilon_1,\epsilon_2,p_1, p_2,  k_1,  k_2$ have
not a simple interpretation in terms of $\vec{\pi}$ and $\vec{r}$ (\ref{f3}). In
fact, there is not a punctual transformation relating both pictures. We shall
comment on this problem more carefully in the last section.

%%%%%%%%%%%%%%%%%%%%%%%%%%%SECTION VI %%%%%%%%%%%%%%%%%%%%%%%%%%%%%%%%
%%%%%%%%%%%%%%%%%%%%%%%%%%%%%%%%%%%%%%%%%%%%%%%%%%%%%%%%%%%%%%%%%%%%%%
\sect{Nonrelativistic anyons in an external electromagnetic field}
\label{galileianyonsfield}

The magnetic limit~\cite{levy} of the Poincar\'e-Maxwell group, that we
call Galilei--Maxwell group~\cite{negro90b}, is the most suitable  to
describe non-relativistic anyons in the presence of external covariant fields.

The Galilei--Maxwell group is a 10--D Lie group, whose
infinitesimal generators are those of $G(2{+}1)$,
$\{H,\,P_1,\,P_2,\,K_1,\,K_2,\,J\}$, together
with $\{ E_1,\,E_2,\, B,\, M\}$.  However, here, we will take into account also
the  central extension characterized by  the nonvanishing commutator
$[K_1,K_2]={\cal K}$, leading to a group  denoted simply 
$GM(2{+}1)$. The nonvanishing commutators of its Lie algebra, 
${\cal GM}(2{+}1)$, are
\begin{equation}\label{commutationsgalileimaxwell}
\begin{array}{lll}
&[E_i, K_j]= -\varepsilon _{ij} B\quad   &[ E_i,J]=-\varepsilon _{ij} E_j   \\[0.25cm]
[H, P_i]=E_{i}  &[H,K_i]=-P_i \quad &[P_i, P_j]= - \varepsilon _{ij} B\\[0.25cm]
[P_i, K_j]=-\delta_{ij} M\quad &[P_i,J]=-\varepsilon _{ij} P_j \quad &[K_i, K_j]=\varepsilon _{ij} {\cal K} \\[0.25cm]
&[K_i,J]=-\varepsilon _{ij} K_j  &\qquad i,j=1,2. 
\end{array}\end{equation}
It is interesting to point out that in a frame where  the electric field $\vec E$
vanishes we recover the commutators  corresponding to a  pure
magnetic Landau system \cite{negro90b,negro02}.
 
On the other hand, it is worthy to note that in order to relate the extended and 
nonextended $GM$ algebras, we can
redefine the basis generators inside the enveloping algebra taking into
account the central character of $M$, 
${\cal K}$ and
$B$. So, we can write
\be\label{changedos}
K'_i= K_i +\lambda \varepsilon_{ij} P_j,\quad
P'_i=P_i + \frac{M}{B} \varepsilon_{ij} E_j
\ee
where $\lambda= (-M+\sqrt{M^2+\k B})/B$, and $M_{e}= \sqrt{M^2+\k B}$ is a kind
of effective mass \cite{duval01}. Then, the new commutators entering $K'_i,P'_i$
are the same as above except that
\be\label{comm}
[K'_i,K'_j] = 0.
\ee
%%%%%%%%%%%%%%%%%%%%%%%%%%
%%%%%%%%%%%%%%%%%%%%%%%%%%
\subsect{Coadjoint orbits}

Let us denote by $(m,\k , \beta , \epsilon_1,\epsilon_2,
h,p_1,p_2,k_1,k_2,j)$ the coordinates of an arbitrary
point of ${\cal G}^*(2{+}1)$ in a basis dual to $(M,\,{\cal
K},\,B,\, E_1,\,E_2,\, H,\, P_1,\, P_2,\, K_1,\, K_2,\,J)$.  The
coadjoint  action of an element
$g=(\theta,\eta,c,\vec{d},b,\vec{a},v,\phi)\in GM(2{+}1)$ on the dual space  
${\cal GM}^*(2{+}1)$ is given by 
\begin{equation}\begin{array}{l}\label{coadjointactiongalileimaxwell}
m' = m , \\[0.25cm]
\vec{\k}' = \vec{\k} , \\[0.25cm]
\vec{\beta }' = \vec{\beta } , \\[0.25cm]
\vec{\epsilon }\:' = \vec{\epsilon }\;^{\phi} - \vec{v}\times \vec{\beta } ,\\[0.25cm]
h'  = h -  \vec{v}\cdot \vec{p}\;^{\phi} + \frac{1}{2}m \vec{v}\;^2
 -\vec{a}\cdot(\vec{\epsilon }~^{\phi}-\vec{v}\times \vec{\beta }),\\[0.25cm]
\vec{p}\:' = \vec{p}\;^{\phi}  -m \vec{v}-b \;(\vec{\epsilon }~^{\phi}
-\;\vec{v}\times \vec{\beta }) - \vec{\beta }\times
\vec{a}  ,\\[0.25cm]
\vec{k}\:' = \vec{k}\;^{\phi}  + m\; \vec{a}+ b\; \vec{p}\;^{\phi} 
-\frac{1}{2} b^2 \vec{\epsilon }\:'
 + \vec{v}\times \vec{\k}
- \vec{\beta }\times\vec{d}+ \frac{1}{2}\vec{a} \times \vec{\beta} ,
\\[0.25cm]
\vec{j}\:' = \vec{j} -  \vec{a}\times \vec{p}\;^{\phi} - \vec{v}\times
\vec{k}\;^{\phi} +
\frac{1}{2} \;\vec{\k}\cdot \vec{v}\;^2 + m \;\vec{a}\times \vec{v}
+\frac{1}{2} \vec{\beta }\cdot \vec{a}\;^2\\[0.15cm]
\hskip1.5cm - \vec{d} \times \vec{\epsilon }\;^{\phi }
+ \frac{1}{2} b\: \vec{\epsilon }\;' 
\times \vec{a}- \vec{\beta } (\vec{d } \cdot \vec{v}) .
\end{array}\end{equation}

Besides $m,\; \k $ and $ C_0\equiv \beta $ we have the following invariants
of the coadjoint action:
\begin{equation}\label{invariantsgalileimaxwell}
C_1=\vec{\beta } \;  \vec{p}\,^2 -2\; m\;h\;\vec{\beta }
+2\; \vec{\beta }\;( \vec{\epsilon }\cdot \vec{k})
 -2\; \beta ^2\; \vec{j}  +\vec{\k} \; \vec{\epsilon }\,^2,\qquad
 C_2= 2\; \vec{\beta }^{\:2}\; h  +2\; \vec{\beta }\cdot( \vec{p}\times 
\vec{\epsilon })+
m\;\vec{\epsilon }\;^2.
\end{equation}
These invariants are the nonrelativistic version of (\ref{pomoc}).
The invariance of  the magnetic field $\vec{\beta }$ may be seen as the consequence
of the invariance of the Lorentz force $\vec{F}=  \vec{\epsilon} + \vec{v} \times
\vec{\beta } $ under (homogeneous)  Galilei transformations. 
A charged particle moving slowly `can see' mainly the magnetic field in
our magnetic limit \cite{levy}:
\be \label{f7}
\frac{\sqrt{\vec{\epsilon }^{\;2}}}{\beta } \ll 1. 
\ee
Let us consider the relativistic invariant $ C_0 = \vec{\epsilon }^{\:2} -\beta^2 $
from (\ref{pomoc}). Using (\ref{f7}) in the $0$-term  approximation
term we obtain the nonrelativistic invariant
\[
C_0= - \beta_0^2.
\]
We can omit the vector symbol because the magnetic field has
only one component, so that
\be
\label{f8}
\beta = \sqrt{\beta_0^2 + \vec{\epsilon }\;^2 } \: \stackrel{\rm(\ref{f7})}{\approx } 
\beta_0 +
 \frac{1}{2}\frac{\vec{\epsilon }\;^2}{\beta_0}.
\ee
By substituting 
$$
h \rightarrow m(1 + \frac{h}{m}),\qquad j \rightarrow -\k(1 - \frac{j}{\k}), 
\qquad
\beta  \rightarrow \beta (1
+ \frac{1}{2}\frac{\vec{\epsilon }\;^2}{\beta^2 })
$$ 
in the other invariants $C_1$ and $C_2$ of (\ref{pomoc}), and
omitting terms of higher order in $ {h}/{m},\; {j}/{\k}$ and  
${\vec{\epsilon }\;^2}/{\beta^2 }$, we obtain their nonrelativistic counterparts 
of (\ref{invariantsgalileimaxwell}), respectively.

 The classification of the coadjoint orbits is displayed in the
Appendix. There are orbits of dimension 6 and 4, but the most important
for us are the 6--D orbits denoted ${\cal O}^{C_1 C_2}_{m \k \beta }$ with
$ \beta \neq 0$.
%%%%%%%%%%%%%%%%%%%%%%%%%
%%%%%%%%%%%%%%%%%%%%%%%%%
\subsection{Symplectic structure}

Each $6$--D orbit ${\cal O}^{C_1 C_2}_{m \k \beta }$ can be covered
with one chart $({\cal O}^{C_1 C_2}_{m \k \beta },\varphi).$ As
coordinates we can choose  $(\epsilon_1,\epsilon_2,p_1,p_2,k_1,k_2)$ since the
Jacobian of the transformation
\[
(m,\k,\beta,\epsilon_1,\epsilon_2,h,p_1,p_2,k_1,k_2,j) \rightarrow   
(m,\k,\beta,C_1,C_2,\epsilon_1,\epsilon_2,p_1,p_2,k_1,k_2)
\]
equals $4 \beta^4\  (\neq 0)$.
The  Poisson tensor $\Lambda$ on the orbit ${\cal O}^{C_1 C_2}_{m
\k \beta }$ is
\begin{equation}\label{lambda6}
\Lambda={m}\;  \frac{\partial}{\partial k_i}\wedge
\frac{\partial}{\partial p_i} + {\k}\;\frac{\partial}{\partial k_1}
\wedge \frac{\partial}{\partial k_2}
- {\beta }\;  \frac{\partial}{\partial p_1} \wedge
\frac{\partial}{\partial p_2} - {\beta }\; \varepsilon^{ij}\;
\frac{\partial}{\partial k_i}\wedge \frac{\partial}{\partial \epsilon _j} .
\end{equation}
The components of $\Lambda$, written in matrix form, are
\begin{equation}
\label{f60}
\Lambda^{ij} = \left(
\begin{array}{cccccc}
0&0&0&0&0& -\beta  \\
0&0&0&0& \beta  &0 \\
0&0&0&-\beta  &-m & 0 \\
0&0&\beta &0&0&-m \\
0&-\beta &m&0&0&\k \\
\beta &0&0&m&-\k&0
\end{array}
\right),
\end{equation}
The determinant of $\Lambda$ is $\beta^6$ ($\neq 0$, for our orbits).
The natural symplectic form on ${\cal O}^{C_1 C_2}_{m \k \beta }$ is
\begin{equation}
\label{symp1}
\omega= \frac{\beta \k + m^2}{\beta ^3} d\epsilon_1 \wedge d\epsilon_2 + 
\frac{m}{\beta^2} d\epsilon_i
\wedge dp_i + \frac{1}{\beta } \varepsilon_{ij}\;dk_i \wedge d\epsilon_j
 + \frac{1}{\beta } dp_1 \wedge dp_2.
\end{equation}
Therefore the coordinates
$(\epsilon_i,p_i,k_i;\; i=1,2)$
 are not canonical since the nonvanishing Poisson brackets are
\be\label{f5}
\{\epsilon_i,k_j\}=  -\varepsilon_{ij} \displaystyle \beta, \qquad
\{p_1,p_2\}=- \beta, \qquad
\{p_i,k_j\}=-\delta_{ij}\; m, \qquad \{k_1,k_2\}=\k.
\end{equation}
Notice in particular that even the coordinates $k_1,k_2$ do not commute.

On the other hand, observe that the above  tensor (\ref{f60})
coincides with (\ref{macierzer}) if we simply substitute $h$ by $m$  and
$j$ by $-\k$. The root of the proposed substitution is the fact
that the Galilei--Maxwell group $GM(2{+}1)$ is the nonrelativistic
limit of the Poincar\'e--Maxwell group $PM(2{+}1).$
We can look at $\k$ as a
nonrelativistic track of the angular momentum $j$ (see also
in this respect the arguments supplied in \cite{jackiw91}).

Using the invariant $C_2$ from (\ref{invariantsgalileimaxwell}) we get a
Hamiltonian linear in momenta $\vec p$ (which is the nonrelativistic
version of (\ref{f})),
\begin{equation}
\label{ff}
h= -\frac{\vec{p} \times \vec{\epsilon }}{ \beta } - \frac{m}{2 \beta ^2}
 \vec{\epsilon }^{\;2} +\frac{C_2}{2 \beta ^2}.
\end{equation}
Observe that by a naive limit $\beta \to 0$ we do not recuperate the free
Hamiltonian (\ref{ha}); in fact 
$\lim_{\beta  \rightarrow 0} h$ is not defined, and the same happens
with the Poisson tensor. 

After simple calculations we obtain the equations
of motion
\begin{equation}\label{el}
\dot{\epsilon _i} =0,\qquad
\dot{p_i} = -\epsilon _i,\qquad
\dot{k_i} = p_i,\qquad
i=1,2.
\end{equation}
Of course, our system includes constant
homogeneous fields $\vec{\epsilon }, \vec{\beta }$ perpendicular to each other as we
already knew from the coadjoint action.
We can also see that $\beta $ does not take part in any of the formulas 
(\ref{el}), so, surprisingly, the equations of motion are not affected by
the magnetic field. Comparing formulas (\ref{el}) with transformation rules
(\ref{coadjointactiongalileimaxwell}) we conclude that these two sets of
equations are compatible if we identify the parameter $b$ with time.
Moreover, the equations of motion are also independent of the parameter
$\k$.

It would be interesting to compare our results with those obtained
in a more standard way following the minimal coupling recipe. For simplicity we
will assume here that the exotic extension $\k$ vanishes. Let us consider a
nonrelativistic particle with a unit negative electric charge moving on a plane in a
constant homogeneous electric
$\vec{\epsilon }=(\epsilon_1,\epsilon_2,0)$ and magnetic 
$\vec{\beta }=(0,0,\beta )$  fields.  A phase space for this
system is a $4$--D symplectic manifold $(M,\tilde{\omega})$, where the
differentiable manifold $M$ is diffeomorphic to $\mathbb{R}^4$ with canonical
coordinates denoted by $\pi_1,\pi_2,r_1,r_2$.
The first pair $\vec{\pi}=(\pi_1,\pi_2,0)$ is interpreted as the generalized
momentum and the second $\vec{r}=(r_1,r_2,0)$ is for the vector of position.
The symplectic form is 
\be \label{ichs} \tilde{\omega}=d\pi_1 \wedge d r_1 + d\pi_2
\wedge dr_2. 
\ee 
The Poisson tensor $\tilde{\Lambda}$ in coordinates $(\pi_1,\pi_2,r_1,r_2)$
takes the natural form
\begin{equation}
\label{ichtensor}
\tilde{\Lambda}^{ij}= \left(\begin{array}{cccc}
0 &0 &-1&0 \\
0&0&0&-1 \\
1 & 0 &0 &0 \\
0 &1 & 0 &0
\end{array} \right).
\end{equation}
Finally, the Hamiltonian is represented by the minimal coupling
expression
\begin{equation}
\label{f9}
H= \frac{1}{2m} \left( \vec{\pi}+\frac{1}{2} \vec{\beta } \times \vec{r}
\right)^2 +  \vec{\epsilon } \cdot \vec{r},
\end{equation}
leading to the motion equations 
\begin{equation}
\label{por0}
\dot{\vec{\pi}}=  -\vec{\epsilon } + \frac{1}{2}\vec{\beta } \times
\dot{\vec{r}},\qquad
\dot{\vec{r}}= \frac{1}{m}\left(
\vec{\pi} + \frac{1}{2}\vec{\beta } \times \vec{r}
\right).
\end{equation}
Integrals of the motion equations (\ref{por0}) are the Hamiltonian (\ref{f9}) and
\bea
& C_1= \vec{\epsilon } \times (\vec{\pi}-\frac{1}{2} \vec{\beta } \times
\vec{r}\:), & \\ 
& C_2=[\vec{\epsilon }\cdot (\vec{\pi}+\frac{1}{2} \vec{\beta }
\times
\vec{r})]^2-2\:(\vec{\epsilon }\cdot \vec{r})[\vec{\beta }\cdot (\vec{\pi}\times
\vec{\epsilon })-m \vec{\epsilon }\:^2]. &
\eea

We can get canonical coordinates from the group coordinates $(\vec{p},\vec{k})$, 
but unfortunately, there is not a point transformation connecting these two
nonrelativistic interacting pictures.

%%%%%%%%%%%%%%%%%%%%%%%%%%%%% CONCLUSIONS %%%%%%%%%%%%%%%%%%%%%
%%%%%%%%%%%%%%%%%%%%%%%%%%%%%%%%%%%%%%%%%%%%%%%%%%%%%%%%%%%%%%%
\section{Concluding remarks}

The symmetry group of a system plus the formalism (Hamiltonian mechanics
on a symplectic manifold) restrict the equations of motion, allow to
define elementary systems, and may lead to interacting systems
compatible with the symmetries. The natural framework to display such
symmetries is the method of coadjoint orbits. In this way we get
a manifold, the invariant symplectic form, and  the
Hamiltonian.

This situation is
quite different for the Hamilton
formulation of mechanics in the phase space. In this frame the same
symplectic manifold $(M,\omega)$ may be used to describe physical
systems with different Hamiltonians. In order to build a Hamiltonian
for interactions one is guided by other principles such as the minimal coupling
rule. However, there is not a canonical way to display the symmetries in this
context.

These two approaches  have significative differences that could
be appreciated  along the examples worked in this paper. For instance, in our
coadjoint orbit scheme, the Hamiltonians obtained for the interacting
cases are linear in
$\vec p$, the equations of motion depend on the electric field $\vec \epsilon$,
while the magnetic field
$\vec{\beta}$ takes part only in the symplectic two--form. These features are in
sharp contrast to the usual interacting Hamiltonian in phase space. 

Another difference is with respect to the role played by the fields.
They are an integral part of the system in the group approach, while in the
phase space they are treated as external parameters. 
The reason is that in our procedure we have considered the fields on the same
foot  as coordinates and momentum. In other words, the fields have been treated
as dynamical fields  instead of external fields, as usual. If we want
to set a complete theory for the whole interacting system (particle + fields) it
is expected that both components should take part of the system at the
same level. 

The group approach also enlighten us how to go from a relativistic to a
nonrelativistic description of the system in a very simple and
natural way. So,  manifolds, symplectic
forms, Hamiltonians, invariants and equations of motion are related through a
contraction procedure.

The price for the simplicity of the group approach is the fact that we
have to use noncanonical and noncommuting coordinates which obscure the
physical interpretation. (However this is usual in the new formulations of planar
physics, see for instance, \cite{horvathy04,lukierski97,bellucci01,lozano01,bak01}). 

To show the explicit relation between  the group approach and
the formulation in phase space is an open problem. This situation of having different descriptions for the same system (one
more appropriate to handle symmetries, the other adapted for an easier
interpretation)  happens also in quantum mechanics.
In this framework symmetries can be described by unitary irreducible
representation of symmetry groups in a representation space related to
the coadjoint orbits as has been shown in Sections 3.3 and 4.3.
On the other side,  quantum mechanical systems are usually described by means
of wavefunctions of the configuration space. The connection between these
two pictures, sometimes is easy (the free case), but when interactions are
included it is more involved.

%%%%%%%%%%%%%%%%%%%%%%%%%%%%%%%%%%%%%%%%%%%%%%%%%%%%%%%%%%%%%%%%%%%%
%%%%%%%%%%%%%%%%%%%%%%%%%%%%%%%%%%%%%%%%%%%%%%%%%%%%%%%%%%%%%%%%%%%%
%%%%%%%%%%%%%%%%%%%%%%% ACKNOWLEDGMENTS %%%%%%%%%%%%%%%%%%%%%%%%%%%%
\section*{Acknowledgments}
This work has been partially supported by DGES of the Ministerio de
Educaci\'on y
Cultura de Espa\~na under Project PB2002--02000, and the
Junta de  Castilla y Le\'on (Spain). One of us (J.T.)  
thanks  the staff of Departamento de
F\'{\i}sica Te\'orica, At\'omica y Nuclear de la Universidad de Valladolid  
for warm hospitality. The authors are indebted Prof. M. Plyushchay for  interesting discussions.
%%%%%%%%%%%%%%%%%%%%%%%%%%%%%%%%%%%%%%%%%%%%%%%%%%%%%%%%%%%%%%%%%%%

 %%%%%%%%%%%%%%%%%%%%%%%%%%%%%%%%%%%%%%%%%%%%%%%%%%%%%%%%%%%%%%%%%%%%%%
%%%%%%%%%%%%%%%%%%%%%%%%%%%%%% BIBLIOGRAPHY %%%%%%%%%%%%%%%%%%%%%%%%%%%%

\pagebreak
%%%%%%%%%%%%%%%%%%%%%%%%%%% APPENDIX %%%%%%%%%%%%%%%%%%
%%%%%%%%%%%%%%%%%%%%%%%%%%%%%%%%%%%%%%%%%%%%%%%%%%%
%%%%%%%%%%%%%%%%%%%%%%%%%%%%%%%%%%%%%%%%%%%%%%%%%%%
\appendix
\renewcommand{\theequation}{A.\arabic{equation}}
\setcounter{equation}{0}

\noindent{\bf\LARGE APPENDICES}
\bigskip

\noindent
{\Large\bf A\hskip0.5cm $G$-invariant Symplectic Structures}\label{symplecticstructures}
\medskip

\noindent
This Appendix contains some basic information about Poisson 
structures on a space ${\cal G}^*$ dual to a Lie algebra ${\cal G}.$ We start
from the definition of the Poisson bracket on some real $n$--dimensional
manifold $M$, then we concentrate on the case when the manifold $M$ is an
orbit ${\cal O}^*$ of the coadjoint action. We prove that ${\cal O}^*$ is
endowed in a canonical way with a symplectic structure.

Let $M$ be a real $n$-dimensional differentiable manifold. The set
of smooth real-valued functions $C^{\infty}(M)$ with a commutative
multiplication constitutes a ring.

The Poisson bracket, 
$\{\cdot, \cdot\}: C^{\infty}(M) \times C^{\infty}(M)
\rightarrow C^{\infty}(M)$, on the manifold $M$ is a bilinear
relation  satisfying the following conditions:
\begin{enumerate}

\item antisymmetry $ \hskip1cm \{f_1,f_2\}= - \{f_2,f_1\} $,

\item  Jacobi's identity $ \hskip0.6cm \{ f_1,\{ f_2, f_3\}\} +
\{ f_2,\{ f_3, f_1\}\} +
\{ f_3,\{ f_1, f_2\}\} =0,$

\item derivation rule \hskip0.9cm  $\{ f_1, f_2 f_3\} = f_2\{ f_1, f_3\} +
f_3\{ f_1, f_2\}, $
\end{enumerate}
for every $f_1,f_2,f_3 \in C^{\infty}(M).$

The two first  properties equip ${\cal C}^{\infty}(M)$ with
the structure of a real Lie algebra. The derivation rule (known also as
the Leibniz identity) and the bilinearity of the Poisson bracket  say
that for every  $f\in {\cal C}^{\infty}(M)$ there exists a vector
field $X_f$ such that 
\[ 
 X_f g=\{ f,g\}, \qquad \forall {g \in {\cal C}^{\infty}(M)}.
\]
Let us cover some open subset $U \subset M$ by a chart
$(U,\varphi)$, such  that $(x_1, \ldots, x_n)$  denotes
the coordinates  of $x\in U$ in this chart.
In a natural basis
$\frac{\partial}{\partial x_i}\equiv
\partial_{x_i} \;\;( i=1, \ldots,  n )$  induced by the chart
we have
 $X_f= \left( X_f\right)^i \partial_{x_i}$, where  Einstein's sum
convention is used. It is easy to check that
\begin{equation}\label{vectorfields}
\left( X_f\right)^i = X_f(x_i)= \{f,x_i\}.
\end{equation}
Using (\ref{vectorfields}) we find that the Poisson bracket
\begin{equation}
\label{dziw0} \{f,g\} = X_f g = \left( X_f\right)^i
\partial_{x_i}g = \{ f,x_i\} \frac{\partial g}{\partial x_i}.
\end{equation}
On the other hand,
\begin{equation}
\label{dziw1} \{ f, x_i\}= -  \{ x_i, f\} = - X_{x_i}(f) =
 - \{x_i, x_j \} \frac{\partial f}{\partial x_j}
=\{ x_j, x_i\} \frac{\partial f}{\partial x_j}.
\end{equation}
Putting (\ref{dziw1}) into (\ref{dziw0}) we finally have
\begin{equation}\label{poissonbracket1}
\{  f,g \}= \frac{\partial f}{\partial x_j} \frac{\partial
g}{\partial x_i} \{ x_j, x_i\}.
\end{equation}
Hence, it is enough to know the Poisson brackets of the coordinate
functions $\{ x_j, x_i\}, \; i,j=1, \ldots, n $, to compute the
Poisson bracket of any pair  of functions. The expression
(\ref{poissonbracket1})  defines a two-contravariant
skew-symmetric tensor $\Lambda$ by
\begin{equation}\label{lambda}
\Lambda(df, dg)=\{  f,g\}.
\end{equation}
So, we
conclude that the correspondence $f \to X_f$ defines a map
$\pi(x) : T_x^{\;*} M\to T_x M$ in every point $x \in M$. The rank of
$\pi(x) $ is called the rank of the Poisson structure in $x$.

The differential equations that determine the integral curves of $X_f$ in
$M$ are
\begin{equation}\label{motionlow}
\frac{d x_i}{d t}=\{  f,x_i\}, \qquad i=1, \ldots, n.
\end{equation}
They look like Hamilton equations being $-f$ the Hamiltonian
function. For this reason the vector fields $X_f$ are called Hamiltonian
vector fields and $f$ is called the Hamiltonian of $X_f$.

Let us consider now the problem of defining a Poisson structure on the
space
 ${\cal G}^*$,  dual to a Lie algebra ${\cal G}$.
For every  smooth function $f\in {\cal C}^\infty({\cal G}^*)$ its
(total) differential
$(d f)_x$ at
$x\in {\cal G}^*$ is a linear mapping
$(d f)_x : T_x {\cal G}^*\to \R$, where $T_x {\cal G}^*$ denotes
the tangent space of ${\cal G}^*$ at the point $x$. The dual space
${\cal G}^*$ is a vector space and  it can be identified with $T_x
{\cal G}^*$. The differential $(D f)_x$ is the  functional over the
tangent space $T_x {\cal G}^*$ and, hence, also over ${\cal G}^*$.
It means that $(D f)_x \in  ({\cal G}^*)^* $ which is isomorphic
to $ {\cal G}$. Thus, to any function  $f\in {\cal C}^\infty({\cal
G}^*)$ we assign $\delta_x f \in {\cal G }$,  in such a way that
for every $y\in {\cal G}^*$
\begin{equation} \label{dual}
\langle y, \delta_x f\rangle =(d\, f)_x(y)=\frac{d}{dt} f(x+ty)|_{t=0}.
\end{equation}
The formula (\ref{dual}) allows us to define the Poisson structure
on $ {\cal G}^*$ by
\begin{equation} \label{dual1}
\{ g, f\}(x)= \langle x, [\delta_x g,\delta_x f]\rangle .
\end{equation}
Defining the functions $\xi _a \in {\cal C}^\infty({\cal G}^*)$, with
$a \in {\cal G}$, by $\xi _a (x)=\langle x, a\rangle$
we obtain  from (\ref{dual}) that
$\delta_x\xi_a=a$. Hence
\begin{equation}\label{poisson1}
\{ \xi _a, \xi _b \}(x)= \langle x, [a,b]\rangle =\xi_{[a,b]}(x) .
\end{equation}
If $\{ a_1,\ldots, a_n\}$ is a basis of ${\cal G}$ then
$\xi_i\equiv \xi_{a_i}, \; i=1, \ldots, n $, can be chosen as  a
set of coordinate functions on ${\cal G}^*$. From
(\ref{poissonbracket1}) and (\ref{dual}) we obtain that the
Poisson structure on ${\cal G}^* $  takes the form
\begin{equation}\label{poisson2}
\{ f  , g \}(x)= \frac{\partial f}{\partial \xi_i}\frac{\partial
g}{\partial
\xi_j} \{ \xi_i  , \xi_j \}=
 \frac{\partial f}{\partial \xi_i}\frac{\partial g}{\partial
\xi_j} \; c^k_{ij} \xi_k ,
\end{equation}
with $c^k_{ij}$ being the structure constants of ${\cal G}$ relative to
its basis $\{ a_1,\ldots, a_n\}$.
Thus, for a set of coordinate functions $\xi_i,\; i=1, \ldots, n$, on
${\cal G}^*$ we have 
\be \label{1q1} 
\{ \xi_i  , \xi_j \}=
c^k_{ij} \xi_k , 
\ee 
so that the Poisson bracket of $\{ \xi_i  , \xi_j
\}$ is a linear function of $\; \xi_k.$

A special case of manifold  with a Poisson structure is the symplectic
manifold.
 A pair $(M,\omega)$ is called a symplectic manifold if
$M$ is a finite-dimensional differentiable manifold and $\omega$  a
nondegenerate $2$-form satisfying the condition
$d\omega =0$.
The nondegeneracy condition of $\omega$ is equivalent to  
the requirement that the rank of  $\hat \omega  : T_x( M)\to T_x^{\;*}( M)$,
defined by
\[
\langle \hat\omega(v),v'\rangle= \omega(v,v'),
\]
be maximal at each point $x\in M$. The map $\hat \omega $  allows us to
define a skewsymmetric contravariant tensor field $\Lambda$ on $M$ by
\begin{equation}\label{lambda1}
\Lambda(\alpha,\beta){=}
\omega (\hat\omega^{-1} (\alpha),\hat\omega^{-1}
(\beta)), \qquad   \alpha, \beta \in  T^*M.
\end{equation}
The tensor $\Lambda$ constitutes the Poisson structure on $M$.
 Thus, when $ \alpha=dg$ and
$\beta = df$ we obtain that
\begin{equation}\label{lambda2}
\{ g, f\}=\Lambda(dg,df)=\omega (\hat\omega^{-1}
(dg),\hat\omega^{-1} (df)).
\end{equation}

Let $G$ be a Lie group and $\cal G$ its Lie algebra. For every element
$g\in
G$ the  inner automorphism  $i_g : G\to G$ defined as
\[
i_g ( g') = g g' g^{-1}
\]
induces  a Lie algebra automorphism $i_{g^*} : {\cal G}\to {\cal G}$
which gives rise to the adjoint representation of $G$ on ${\cal G}$ by
${\rm Ad}_g=\exp{i_{g^*}}$. The coadjoint representation of $G$ on
${\cal G}^*$ is now given  by
\begin{equation} \label{coadjoint}
\langle {\rm CoAd}_g(u),a\rangle= \langle u, {\rm Ad}_{g^{-1}}(a)\rangle ,
\qquad u\in {\cal G}^*,\;\; a\in {\cal G}.
\end{equation}
Each orbit $\cal O^*$ of the coadjoint action  is a symplectic
submanifold of the Poisson manifold ${\cal G}^*$, and it is
endowed in a canonical way with a symplectic structure
characterized by the two-form (Kirillov--Kostant-Souriau theorem)
\begin{equation} \label{2formcoadjoint}
\omega (X_a,X_b)= \langle u, [a,b]\rangle , \qquad u\in {\cal
O}^*,\;\; a, b\in {\cal G},
\end{equation}
where $X_a$ is the fundamental vector field associated with the coadjoint
action
\begin{equation} \label{fundamentalfield}
(X_a f)(u)= \frac{d}{dt} f\left({\rm CoAd}_{e^{-ta}} u
\right)\left|_{t=0} \right. .
\end{equation}

The Poisson structure on ${\cal O}^*$, as a submanifold of ${\cal
G}^*$, defined through the symplectic 2-form $\omega$
(\ref{2formcoadjoint}) coincides with that induced by the Poisson
structure (\ref{poisson2})   on ${\cal G}^*$.

\newpage
{\Large\bf B\hskip0.5cm Coadjoint orbits classification for ${\cal GM}^*(2{+}1)$}
\label{class}

\begin{tabular}{| c  | c |c|}  \hline
Constraints & Dim & Invariants \\  \hline $\beta  \neq 0, \; m \neq 0,
\; \k \neq 0 $ & 6 & $  C_1= \frac{1}{2} \vec{\varepsilon }^{\:2}\cdot
\vec{\k} + (\vec{\varepsilon }\cdot \vec{k})\cdot
                                                 \vec{\beta }+\frac{1}{2}\vec{p}^{\:2} \cdot \vec{\beta } - m h \vec{\beta } -(\vec{j}\cdot
                                                  \vec{\beta }) \cdot \vec{\beta } $   \\
 & & \\
                                        &   & $  C_2= \vec{\beta }^2 h +   (\vec{p} \times \vec{\varepsilon }\:) \cdot \vec{\beta } +
                                                  \frac{m}{2}\vec{\varepsilon }^{\:2}  $ \\
& & \\ \hline
& & \\
$ \beta  = 0, \; m \neq 0, \; \k \neq 0 $ & 6  & $
C_1=-\frac{m}{2}\vec{p}^{\:2} + m^2 h -m \vec{\varepsilon } \cdot \vec{k} +
(\vec{p}
                                                \times \vec{\varepsilon }\:) \cdot  \vec{k} $ \\
& & \\
                                        &   & $ C_2= \vec{\varepsilon }^{\:2} $ \\
& & \\ \hline
& & \\
$ m = 0, \; \beta  \neq 0, \; \k \neq 0 $ & 6 & $ C_1= \frac{1}{2}
\vec{\varepsilon }^{\:2}\cdot \vec{\k} + (\vec{\varepsilon }\cdot \vec{k})\cdot
                                               \vec{\beta }+ \frac{1}{2}\vec{p}^{\:2} \cdot \vec{\beta }  - (\vec{j}\cdot
                                                 \vec{\beta }) \cdot \vec{\beta } $ \\
& & \\
 &   & $ C_2= h \cdot \vec{\beta } +  (\vec{p} \times \vec{\varepsilon }\:) $ \\
& & \\ \hline
& & \\
$ \k= 0, \; \beta  \neq 0, \; m \neq 0 $ & 6  & $ C_1=   (\vec{\varepsilon }\cdot
\vec{k})\cdot \vec{\beta }+\frac{1}{2}\vec{p}^{\:2} \cdot \vec{\beta } - m h \cdot  \vec{\beta } -
                                                (\vec{j}\cdot
                                                \vec{\beta }\:) \cdot \vec{\beta } $ \\
& & \\
 &   & $ C_2= \vec{\beta }^{\:2} h +  (\vec{p} \times \vec{\varepsilon }\:) \cdot \vec{\beta } +
\frac{m}{2}\vec{\varepsilon }^{\:2} $ \\
& & \\ \hline
& & \\
$ \beta =m= 0, \;  \k \neq 0 $ & 6 & $C_1=\vec{p} \times \vec{\varepsilon } $ \\
& & \\
& & $ C_2=\vec{\varepsilon }^{\:2} $ \\
& & \\ \hline
& & \\
$ \beta =\k= 0, \;  m \neq 0 $ & 6 & $ C_1= - \frac{1}{2}\vec{p}^{\:2} + m  h - \vec{\varepsilon } \cdot \vec{k} $ \\
& & \\
& &  $C_2=\vec{\varepsilon }^{\:2} $ \\
& & \\ \hline
& & \\
$ m=\k= 0, \;  \beta  \neq 0 $ & 6 & $ C_1=- \frac{1}{2}\vec{p}^{\:2} -\vec{\varepsilon } \cdot \vec{k} + \vec{\beta } \cdot \vec{j} $ \\
& & \\
& & $ C_2=h \cdot \vec{\beta } + \vec{p} \times \vec{\varepsilon } $ \\
& & \\ \hline
& & \\
$ \beta =m=\k= 0 $ & 4  & $ C_1= \frac{1}{2}\vec{p}^{\:2} + \vec{\varepsilon  } \cdot \vec{k} $ \\
& & \\
& & $ C_2= \vec{p} \times \vec{\varepsilon } $ \\
& & \\
& & $C_3=\vec{\varepsilon }^{\:2} $ \\
& & \\
& & $ C_4= (\vec{p} \cdot \vec{\varepsilon }\:) \cdot (\vec{p} \times \vec{\varepsilon }\:)
+ \vec{\varepsilon }^{\:2}
     \cdot  (\vec{k} \times \vec{\varepsilon }\:) $ \\
 \hline
\end{tabular}
%%%%%%%%%%%%%%%%%%%%%%%%%%%%%%%%%%%%%%%%%%%%%%%%%%%%%%%%%%%%%%%%%
\end{document}